\documentclass[prd,preprint,superscriptaddress,showpacs]{revtex4}
\usepackage{graphicx}
\usepackage{latexsym}
\usepackage{amsmath}
\usepackage{subfigure}

\begin{document}

\title{Constraining the interacting dark energy models from weak gravity conjecture and recent observations}
\author{Ximing Chen}
 \email{chenxm@cqupt.edu.cn}
 \affiliation{College of Mathematics and Physics, Chongqing University of Posts and Telecommunications,
Chongqing 400065, China}
\author{Bin Wang}
\email{wang_b@sjtu.edu.cn}
\affiliation{College of Mathematics and Physics, Chongqing University of Posts and Telecommunications,
Chongqing 400065, China}
\affiliation{Department of Physics, Shanghai Jiao Tong University, Shanghai 200240, China}
\author{Nana Pan}
\email{pannn@cqupt.edu.cn}
 \affiliation{College of Mathematics and Physics, Chongqing University of Posts and Telecommunications,
Chongqing 400065, China}
\author{Yungui Gong}
\email{gongyg@cqupt.edu.cn}
 \affiliation{College of Mathematics and Physics, Chongqing University of Posts and Telecommunications,
Chongqing 400065, China}

\begin{abstract}
We examine the effectiveness of the weak gravity conjecture in constraining the dark energy by comparing with observations. For general dark
energy models with plausible phenomenological interactions between dark sectors, we find that although the weak gravity conjecture can constrain
the dark energy, the constraint is looser than that from the observations.
\end{abstract}

\preprint{arXiv: 1008.3455}

\maketitle
\section{Introduction}
Our universe is believed undergoing an accelerated expansion driven by a yet unknown dark energy (DE) \cite{zhu2, zhu3}. The leading interpretation of such DE is
the vacuum energy with equation of state (EoS) $w = -1$. Although this interpretation is consistent with observational data, at the fundamental
level it fails to be convincing, since its energy density falls far below the value predicted by any sensible quantum field theory, and it
unavoidably leads to the coincidence problem \cite{zhu4}. It is expected that the string theory can provide the resolution to these problems and disclose the
nature of DE. In string theory there are vast amount of landscape vacua which can be constructed and described by the low-energy effective field
theories. However a large number of these semi-classically consistent effective field theories are found inconsistent at quantum
level \cite{zhu6, zhu7}. Recently it was argued that the weak gravity conjecture can be used to rule out the effective field theory which is not
consistent with the full quantum theory \cite{zhu8}. This conjecture was generalized to asymptotically dS/AdS background \cite{zhu9}, leading to an
upper bound on the cosmological constant, and applied to inflationary cosmology \cite{zhu10}. For the DE problem, if we consider that
our universe is one of the vast landscape of vacua, then employing a low energy effective field theory to describe the vacuum energy, the
variation experienced by the DE scalar field from any redshift within the classical expansion era till now should not exceed Planck's
mass \cite{zhu11},
\begin{equation}
\label{weakgravconj}
\frac{\Delta\phi(z_m)}{M_p} = \int \frac{\dot\phi}{M_p} dt
=\int_0^{z_m}\frac{\sqrt{3\mid[1+w(z)]\Omega_{\phi}(z)\mid}}{1+z} dz<1,
\end{equation}
where $z_m$ is the highest redshift to the last scattering surface, $H$ is the Hubble parameter and $M_p^{-2}$ is the Planck mass.

The bound (1) is the realization of the weak gravity conjecture on the scalar field. It was argued that this bound provides a theoretical
constraint on the DE EoS \cite{pavon1,pavon2,Chen}. Different from the observational constraints such as the supernova Ia (SNIa), cosmic microwave
background (CMB), baryon acoustic oscillations (BAO) etc, the theoretical
bound is a natural condition and does not impose any prior upon parameter space. It was claimed that this theoretical constraint is more
stringent than the constraint from observations \cite{pavon1}. The theoretical condition has been further employed to constrain the
Chaplygin-Gas-Type DE model \cite{zhu}, the holographic and the agegraphic DE models \cite{zhangxin} etc. More recently this theoretical bound was
used to constrain the interacting DE models \cite{pavon}. Assuming that DE interacts with dark matter (DM) through a small coupling, we can alleviate the cosmic
coincidence problem \cite{wang}. Encouragingly it was found that this interaction is allowed by observations including the universe expansion
history \cite{wang1,guozk} and galaxy scale observations \cite{wang2} etc. For the time being, since we know neither the nature of DE nor DM, we
can only guess plausible forms of the interaction between dark sectors on phenomenological bases. In \cite{pavon}, taking the interaction between
DE and DM proportional to the energy density of DE or DM, the constraint on the DE EoS was obtained by using the theoretical bound. In order to
further examine the validity of the theoretical bound and its effectiveness on disclosing the nature of DE, in this work we generalize the
discussion in \cite{pavon} to constrain the DE EoS when DE and DM are interacting with other phenomenological couplings. Instead of just
examining the DE EoS from the theoretical bound as carried out in \cite{pavon}, we will combine our result with the observational constraints.
This can help us further examine the effectiveness of the theoretical constraint on the nature of the DE.

In the presence of the interaction between DE and DM, the continuity equations for the DE and DM are
\begin{eqnarray}\label{eom1}
\dot{\rho}_m+3H\rho_m=-Q,
\end{eqnarray}
\begin{eqnarray}\label{eom2}
\dot{\rho}_\phi+3H(\rho_\phi+p_\phi)=Q,
\end{eqnarray}
where $Q$ indicates the coupling between dark sectors and in our discussion it will be specified with two phenomenological forms, such as
$Q_1=\alpha\kappa^{2} H^{-1}\rho_m\rho_{\phi}$ and $Q_2=\alpha\kappa^{2n} H^{3-2n} \rho_m^n$, respectively. Here $\alpha$ is the strength of the
interaction and $\kappa^2=8\pi/M^2_{p}$. The dynamics and observational constraints on these two interaction forms have been discussed in \cite{he, he14,Chen:2008pz,wang3}.
When $Q>0$, we have the energy transfer from DM to DE.

In the background flat FRW universe, the Friedmann equations used to govern the evolution of the universe have the form
\begin{equation}\label{FR1}
H^{2}=\frac{\kappa^{2}}{3}(\rho_{m}+\rho_{\phi}),
\end{equation}
\begin{equation}\label{hubeq}
\frac{\dot{H}}{H^2}=-\frac{3}{2}[1+w(z)\Omega_{\phi}],
\end{equation}
where we have neglected the baryonic matter and set $\Omega_{m}=\kappa^2\rho_{m}/(3H^2)$ and $\Omega_{\phi}=\kappa^2\rho_{\phi}/(3H^2)$. Defining
$1+z=a_0/a$ and substituting $\rho_{\phi}=3\kappa^{-2}H^2\Omega_{\phi}$ into (\ref{eom2}), we can obtain the evolution equation of
$\Omega_{\phi}$ along the redshift $z$ by combining (\ref{FR1}) and (\ref{hubeq}),
\begin{equation}\label{dotz2}
\frac{d\Omega_{\phi}}{dz}=w(z){\Omega_{\phi}(1-\Omega_{\phi})\frac{3}{1+z}}
-\frac{\kappa^2Q}{3H^{3}}\frac{1}{1+z}.
\end{equation}
Substituting the phenomenological interaction form $Q$ and the EoS of DE, we can employ the weak gravity bound (1) to examine the property of DE.
In the following discussion, we will consider the constant DE EoS $w_1(z)=w_{0}=const$ and a time-dependent DE EoS in the form
$w_2(z)=w_0\exp[z/(1+z)]/(1+z)$ \cite{gong051par}.

\section{Constraints from weak gravity conjecture and observations}

In order to examine the validity and effectiveness of the theoretical bound, we will compare its derived constraint on DE with the observational
constraint. The dimensionless Hubble parameter is expressed as
\begin{equation}\label{hubeq2}
E^2(z)=\frac{H^{2}}{H^{2}_0}=\frac{\kappa^2}{3H_0^{2}}(\rho_m+\rho_{\phi})
=\frac{\rho_{m}}{\rho_{m0}}\Omega_{mo}+\frac{\rho_{\phi}}{\rho_{\phi0}}\Omega_{\phi0},
\end{equation}
where $\Omega_{m0}\equiv\kappa^2\rho_{m0}/3H^2_{0}$ and $\Omega_{\phi0}\equiv\kappa^2\rho_{\phi0}/3H^2_{0}$.
For the observational data, we take SNIa, BAO and CMB data. For the SNIa data, we first use the recent Union2
compilation of 557 SNIa which employs the SALT2 light curve fitter \cite{union2}, and then we use the Constitution
compilation of 397 SNIa which employs the SALT light curve fitter \cite{constitution}, for the check of systematics.
For the Union2 SNIa data, we add the covariant matrix which includes the systematical errors \cite{union2}. For the BAO measurement
from the Sloan digital sky survey (SDSS), we follow \cite{bao1} to use the ratio of angular distance $d_A(z)$ and the dilation scale
$D_v(z)$ which combines the BAO measurements at two different redshifts $z=0.2$ and $z=0.35$ \cite{bao2}, with the angular scale of
the sound horizon at recombination $l_A$ measured from CMB by WMAP5 \cite{wmap5}. The redshifts of drag epoch $z_d\approx 1020$ and
recombination $z_*\approx 1090$ were
chosen from WMAP5 data \cite{wmap5}. In addition, we also use the CMB shift parameter $R$ as measured from WMAP5 \cite{yun06,wmap5}.
For the observation constraint we
will constrain the interacting DE model by minimizing the quantity
\begin{eqnarray}
\label{chi}
\lefteqn{\chi^2 = \sum_{i,j}[\mu_{obs}(z_i)-\mu(z_i)]Cov^{-1}_{sn}(\mu_i,\mu_j)[\mu_{obs}(z_j)-\mu(z_j)]+{}}  \nonumber\\
&&{}\left(\frac{\frac{d_A(z_*)}{D_v(0.2)})-17.55}{0.65}\right)^2+\left(\frac{\frac{d_A(z_*)}{D_v(0.35)})-10.10}{0.38}\right)^2
+\left(\frac{R-1.710}{0.019}\right)^2,
\end{eqnarray}
where the distance modulus $\mu(z)=5\log_{10}[d_L(z)/{\rm Mpc}]+25$, the luminosity distance $d_L(z)$ of supernovae is given by
\begin{equation}
\label{lumdis}
d_L(z)=\frac{1+z}{H_0} \int_0^z
\frac{dx}{E(x)},
\end{equation}
the angular distance $d_A(z)=d_L(z)/(1+z)^2$, the dilation scale
\begin{equation}
\label{dvdef}
D_V(z)=\left[\frac{d_L^2(z)}{(1+z)^2}\frac{z}{H(z)}\right]^{1/3},
\end{equation}
and the CMB shift parameter is \cite{yun06,wmap5}
\begin{equation}
\label{shift} R=\sqrt{\Omega_{m0}}\int_0^{z_*} \frac{dx}{E(x)}=1.710\pm0.019.
\end{equation}

In the following we report the constraint obtained by the weak gravity conjecture and examine its validity
and effectiveness by comparing it with
the observational constraints for our models with selected interaction forms and different DE EoS.

\subsection{Interaction form 1}

We first concentrate on the phenomenological coupling involving both dark sectors, which can be adopted as the
product of the densities of DE and DM in the form $Q_1=\alpha\kappa^{2} H^{-1}\rho_m\rho_{\phi}$. We will choose two different DE EoS in our
discussion, the constant EoS $w_1(z)=w$ and the time-varying EoS $w_2(z)=w_0\exp[z/(1+z)]/(1+z)$.

For the constant DE EoS, we can easily get the evolution of $\Omega_{\phi}$ along the redshift $z$
\begin{equation}\label{dotz4}
\frac{d\Omega_{\phi}}{dz}=\frac{3(1-\Omega_{\phi})}{1+z}(w-\alpha)\Omega_{\phi}.
\end{equation}
Substituting the integration of (\ref{dotz4}) into  (\ref{weakgravconj}), we can get the result $\Delta\phi(z_m)/M_p$. Here we have three free
parameters, ($\Omega_{m0}$, $w$, $\alpha$). To satisfy the weak gravity conjecture, we require $\Delta\phi(z_m)/M_p<1$. Taking $z_m=1090$, we
have the theoretical constraints $0.03\le\Omega_{m0}\le0.40$ and $-1.20\le{w}\le{-0.65}$ by fixing $\alpha_c=0.04926$. Fixing
$\Omega_{m0c}=0.27699$, the weak gravity condition gives  $-0.61\le{\alpha}\le 1.00$ and $-1.20\le {w}\le {-0.23}$. Fixing $w_{c}=-1.0588$, we have
the theoretical constraints  $0.02\le{\Omega_{m0}}\le0.4$ and $-0.80\le{\alpha}\le{-0.32}$. Here $\alpha_c$, $\Omega_{m0c}$, $w_c$ are the central values
from the observational constraints below. We see that the theoretical allowed ranges of the parameters are large, which shows that the
constraint given by the weak gravity conjecture is loose in this case.

To compare with the observational constraint, we write out the evolving equations of $\rho_{m}/\rho_{m0}$ and $\rho_{\phi}/\rho_{\phi 0}$
\begin{equation}\label{rhozeq1}
\frac{d\rho_m/\rho_{m0}}{dz}=\frac{3}{1+z}\frac{\rho_{m}}{\rho_{m0}}
+\frac{3\alpha}{1+z}\frac{1-\Omega_{m0}}{E^2(z)}\frac{\rho_{m}}{\rho_{m0}}\frac{\rho_{\phi}}{\rho_{\phi0}},
\end{equation}
\begin{equation}\label{rhozeq2}
\frac{d\rho_{\phi}/\rho_{\phi0}}{dz}=\frac{3[1+w_0]}{1+z}\frac{\rho_{\phi}}{\rho_{\phi0}}
-\frac{3\alpha}{1+z}\frac{\Omega_{m0}}{E^2(z)}\frac{\rho_{m}}{\rho_{m0}}\frac{\rho_{\phi}}{\rho_{\phi0}}.
\end{equation}
Using the combination of the recent Union2 SNIa data \cite{union2}, the BAO data  \cite{bao1} and the CMB shift data from WMAP5 \cite{wmap5},
we have the best-fit values
$\Omega_{m0}= 0.277^{+0.073}_{-0.055}$, $w=-1.059^{+0.368}_{-0.482}$ and $\alpha= 0.049^{+0.177}_{-0.107}$ with
$\chi^{2}=547.05$ at the $3\sigma$ confidence level. The interaction strength is constrained within the range $ -0.058\le{\alpha}\le0.226$ at
the $3\sigma$ confidence level. For comparison and systematic check, we also replace the Union2 SNIa
data by the Constitution SNIa data \cite{constitution}, to examine this model too. The best-fit results are $\Omega_{m0}=
0.281^{+0.067}_{-0.062}$, $w=-0.988^{+0.213}_{-0.304}$ and $\alpha= 0.036^{+0.197}_{-0.110}$ with $\chi^{2}=466.39$ at the $3\sigma$
confidence level. The strength of the interaction is in the range $-0.074 \le{\alpha}\le 0.233$ at the $3\sigma$ confidence level.
So the constraint by the Union2 SNIa data is better due to more SNIa data.

In Fig. 1 we plot the results from the theoretical and observational constraints by choosing the best fitted values
$\alpha_c$, $\Omega_{m0c}$, and $w_c$,
respectively. It is easy to see that the allowed parameter space by the theoretical constraint is much bigger than that in the $3\sigma$
observational constraint. Thus for the interaction proportional to the product of DE and DM energy densities, when the DE EoS is a constant, the
weak gravity conjecture is less stringent than the observational constraint. We also plot $\Delta\phi/M_p$ versus $z_m$ in Fig.
\ref{fig2:subfig:a} for different $\Omega_{m0}$ , $w$ and $\alpha$ for the interacting dark energy form $Q_1$ and the weak gravity conjecture
is always respected.

For the time-dependent DE EoS in the form $ w_2(z)=w_0\exp[z/(1+z)]/(1+z)$, we examine again the effectiveness of the weak gravity
conjecture. The ratio of the variation experienced by the DE scalar field and the Planck's mass reads
\begin{equation}
\label{weakgravconj2}
\frac{\Delta\phi(z_m)}{M_p}=
\int_0^{z_m}\frac{\sqrt{3\mid[1+w_0\exp(z/(1+z))/(1+z)]\Omega_{\phi}(z)\mid}}{1+z} dz,
\end{equation}
where the evolution of the DE is described by
\begin{equation}\label{dotz4b}
\frac{d\Omega_{\phi}}{dz}=\frac{3(1-\Omega_{\phi})}{1+z}[w_0\exp(z/(1+z))/(1+z)-\alpha]\Omega_{\phi}.
\end{equation}
Substituting the integration of (\ref{dotz4b}) into (\ref{weakgravconj2}), we can get the result $\Delta\phi(z_m)/M_p$. Here again we have three
free parameters ($\Omega_{m0}$, $w_0$, $\alpha$). Requiring $\Delta\phi(z_m)/M_p<1$, where $z_m=1090$, we have the theoretical constraints on the
parameter space. Fixing $\alpha_c=0.02194$, we have  $0.07\le\Omega_{m0}\le0.69$ and $-4.0\le{w_0}\le{-0.87}$ to satisfy
$\Delta\phi(1090)/M_p<1$. Fixing $\Omega_{m0c}=0.27592$, the weak gravity conjecture tells us that $-0.44\le{\alpha}\le1.00$ and
$-3.0\le{w_0}\le{-0.28}$. Furthermore fixing $w_{0c}=-1.0700$, we have  $0.01\le{\Omega_{m0}}\le0.69$ and $-0.09\le{\alpha}\le{-0.65}$. Here $\alpha_c$,
$\Omega_{m0c}$, $w_{0c}$ are the central values from the observational constraint below.

To check the effectiveness of the theoretical constraint, we also  fit the Union2 SNIa+BAO+$R$  data for the model with varying DE EoS, we get
$\Omega_{m0}= 0.276^{+0.073}_{-0.055}$, $w_0=-1.070^{+0.373}_{-0.482}$ and $\alpha= 0.022^{+0.183}_{-0.127}$ with
$\chi^{2}=546.94$ at the $3\sigma$ confidence level. By fitting to the Constitution SNIa+BAO+$R$ data sets, the joint constraints are $\Omega_{m0}=
0.280^{+0.068}_{-0.062}$, $w_0=-0.995^{+0.214}_{-0.303}$ and $\alpha= 0.003^{+0.203}_{-0.122}$ with $\chi^{2}=466.41$ at the $3\sigma$
confidence level. The constraint by the Union2 SNIa data is a little better than that by the Constitution SNIa data.

In Fig. 3 we plot the results on the parameter space obtained from the theoretical constraint and the observational constraint by fixing the
central values $\alpha_c$, $\Omega_{m0c}$, and $w_{0c}$, respectively. Comparing with the result for the constant DE EoS, we see that the allowed region of
the parameter space from the weak gravity conjecture is significantly reduced. If we combine the theoretical constraint with the observational
constraint, we can have the stringent constraint on the parameter space.
\begin{figure}
  \centering
  \subfigure[]{
    \label{fig1:subfig:a}
    \includegraphics[width=6cm]{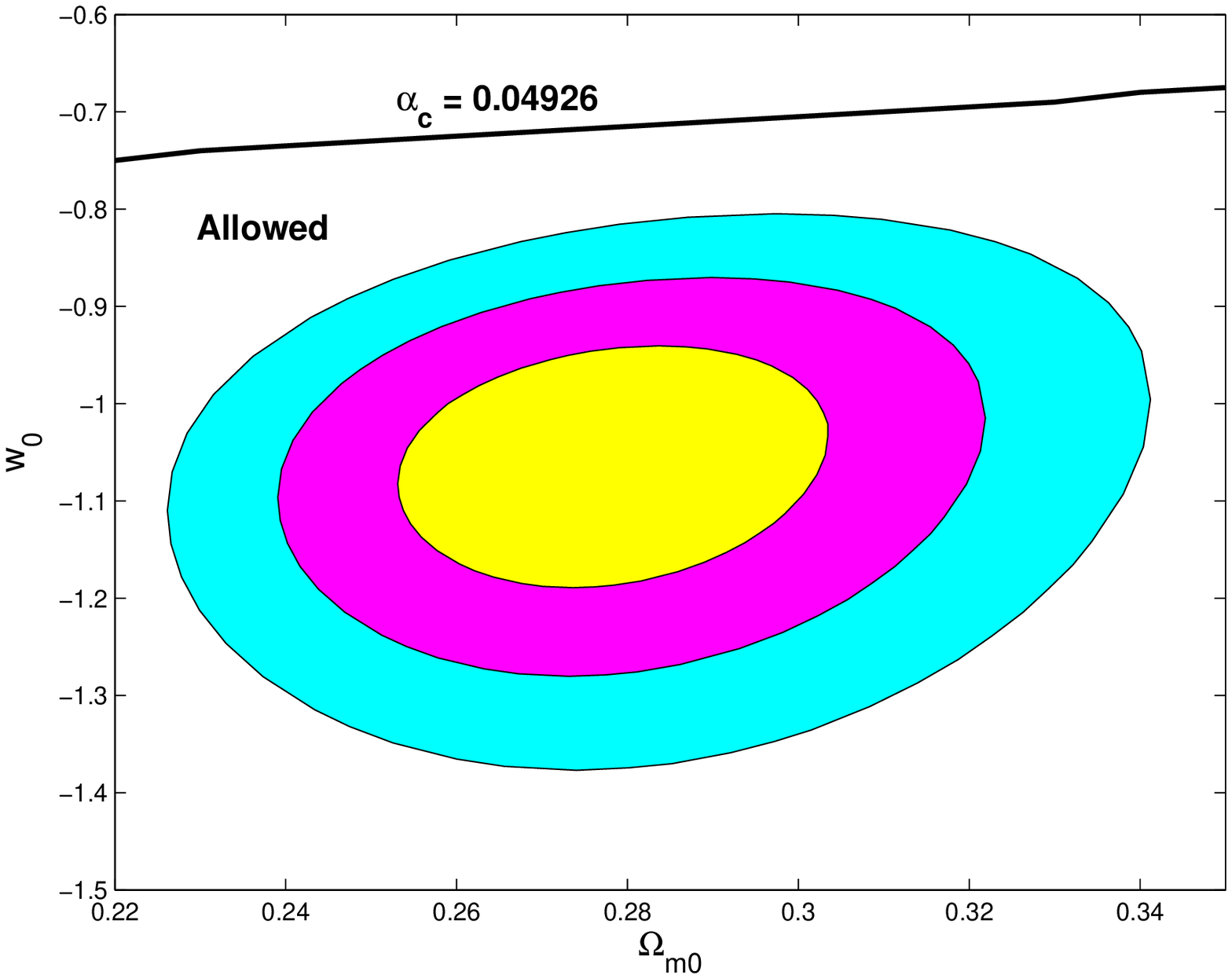}}
  \hspace{1in}
  \subfigure[]{
    \label{fig1:subfig:b}
    \includegraphics[width=6cm]{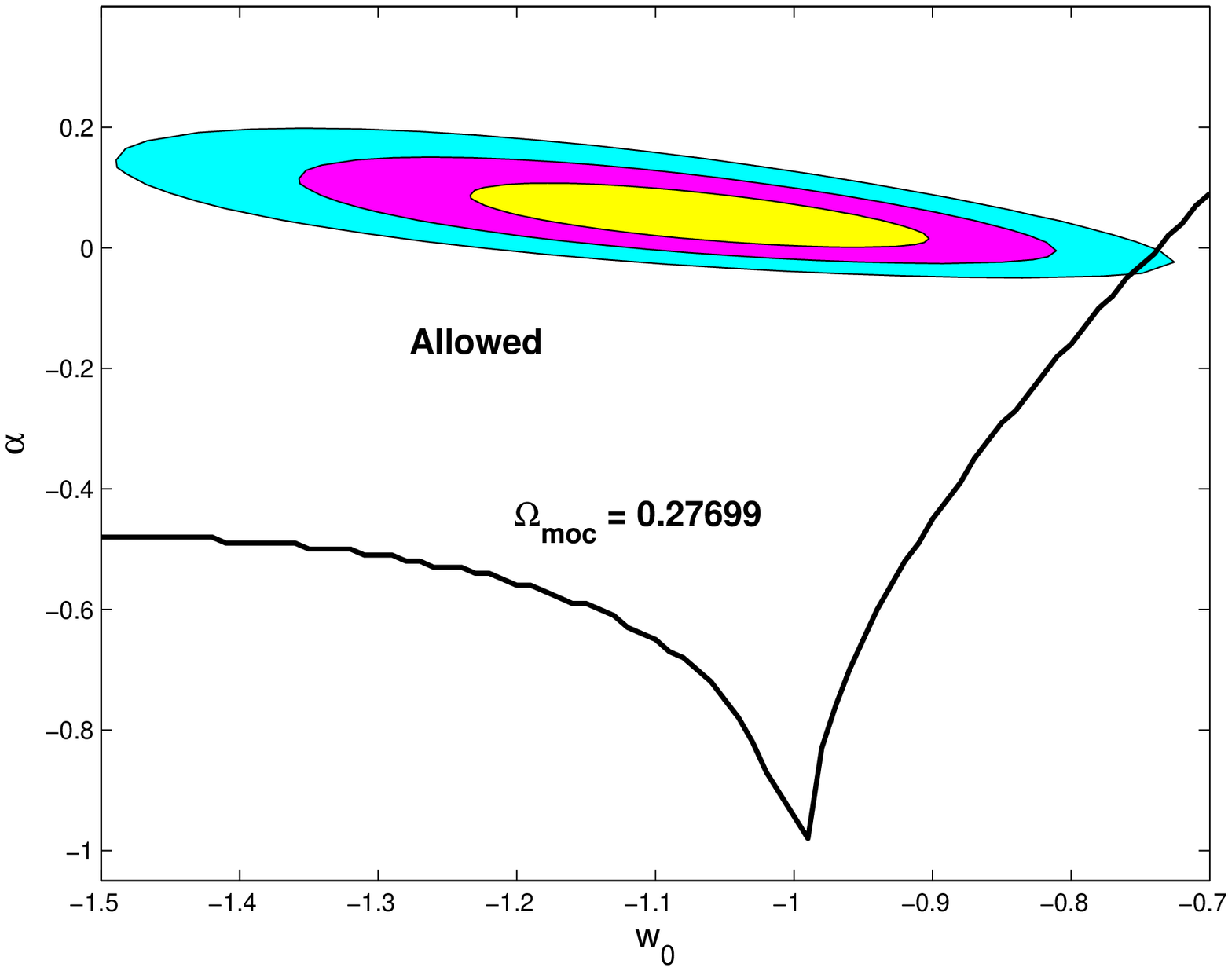}}
     \hspace{1in}
      \subfigure[]{
    \label{fig1:subfig:c}
    \includegraphics[width=6cm]{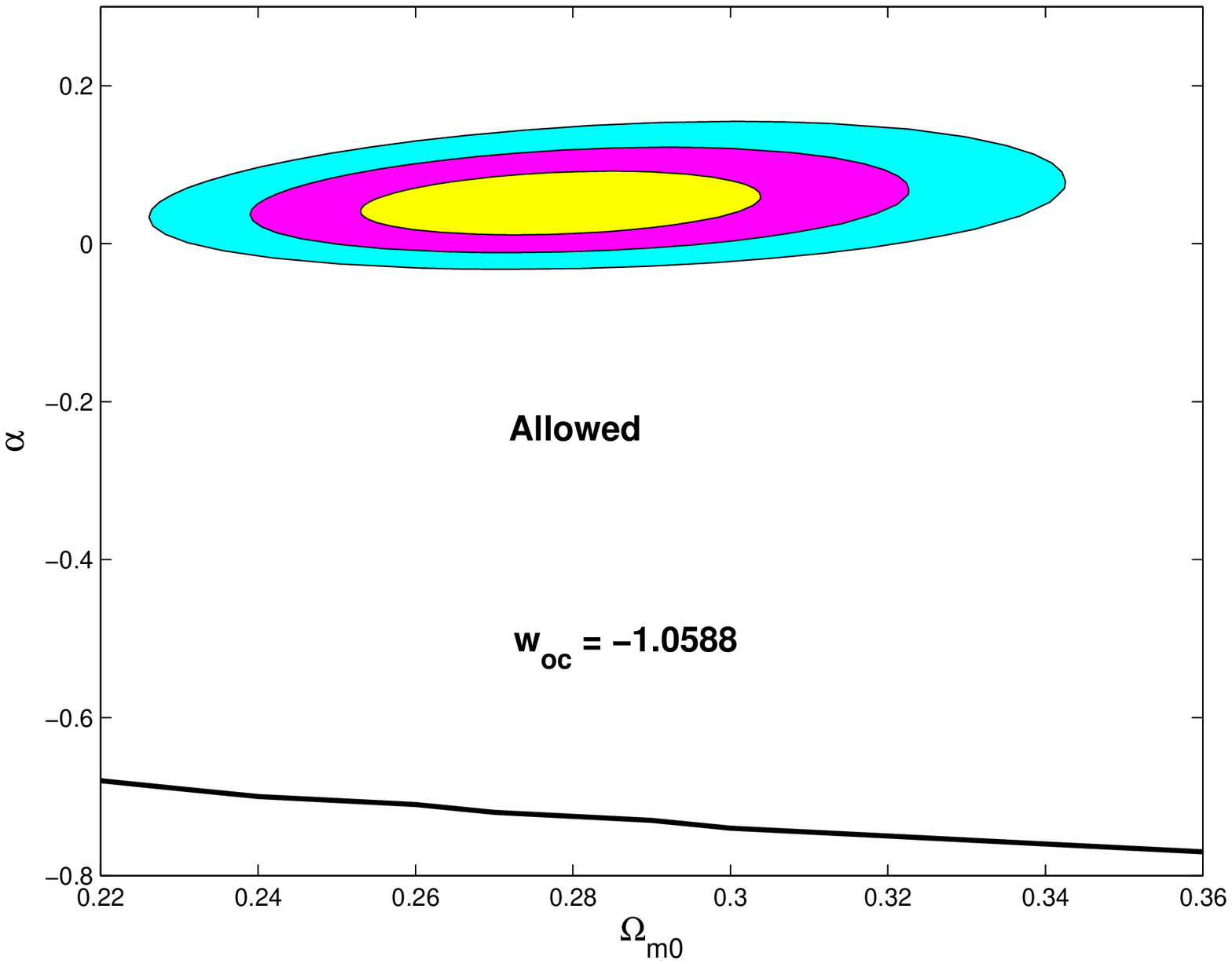}}
     \hspace{1in}
\caption{Contours on the parameter space from the observational fitting result by using the Union2 SNIa, the BAO measurement from SDSS and the
CMB shift parameter from WMAP5 for the interaction between dark sectors with the form $Q_1$ and constant DE EoS $w_1(z)=const$. The solid line
indicates the constraint from the theoretical condition. ``allowed" indicates the region permitted by the weak gravity conjecture. In (a) we
choose the  $\alpha_c=0.04926$. In (b) we take $\Omega_{m0c}=0.27699$. In (c), $w_{0c}=-1.0588$.}
  \label{fig1}
\end{figure}

\begin{figure}
  \centering
    \subfigure[$Q_1=\alpha\kappa^{2} H^{-1}\rho_m\rho_{\phi}$]{
    \label{fig2:subfig:a}
    \includegraphics[width=6cm]{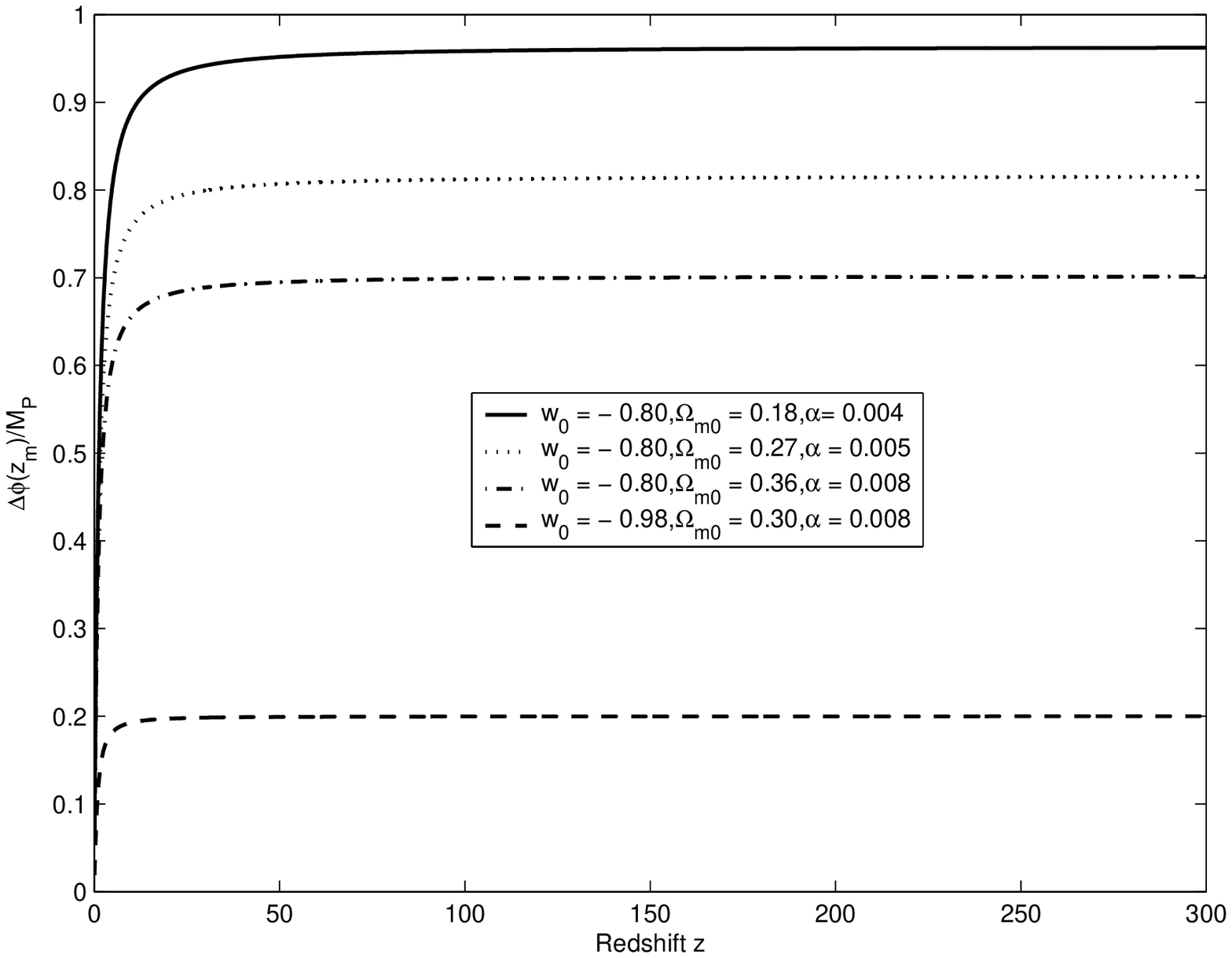}}
  \hspace{1in}
    \subfigure[$Q_2=\alpha \kappa^{2} H^{-1} \rho_m^2$]{
    \label{fig2:subfig:b}
    \includegraphics[width=6cm]{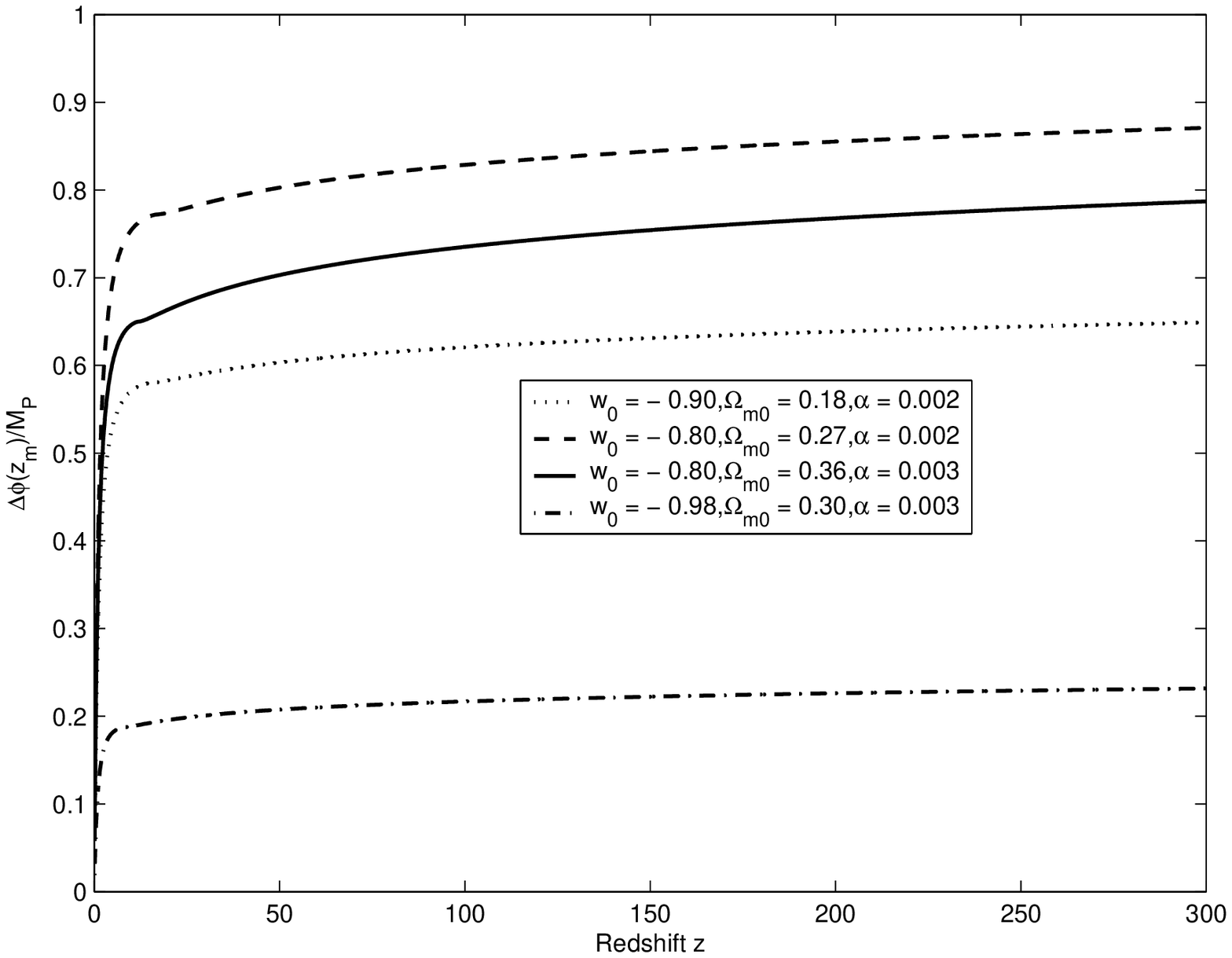}}
     \hspace{1in}
\caption{$\Delta\phi(z_m)/M_p$ versus the maximum redshift $z_m$ for different $\Omega_{m0}$ , $w_0$ and $\alpha$ for two interacting dark energy
forms $Q_1$ and $Q_2$ with $w_1(z)=w_0=const$.} \label{fig2}
\end{figure}

\begin{figure}
  \centering
  \subfigure[]{
    \label{fig3:subfig:a}
    \includegraphics[width=6cm]{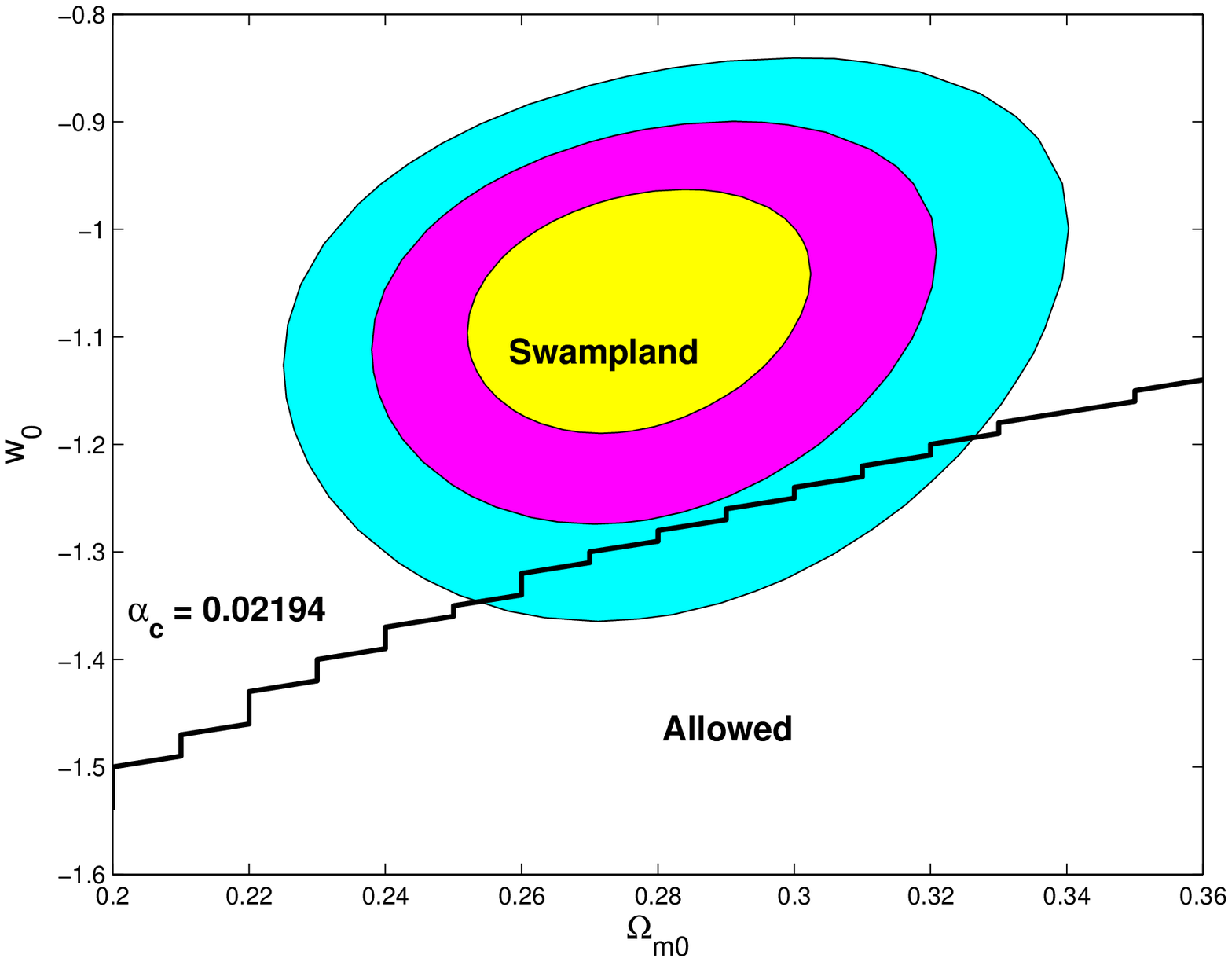}}
  \hspace{1in}
  \subfigure[]{
    \label{fig3:subfig:b}
    \includegraphics[width=6cm]{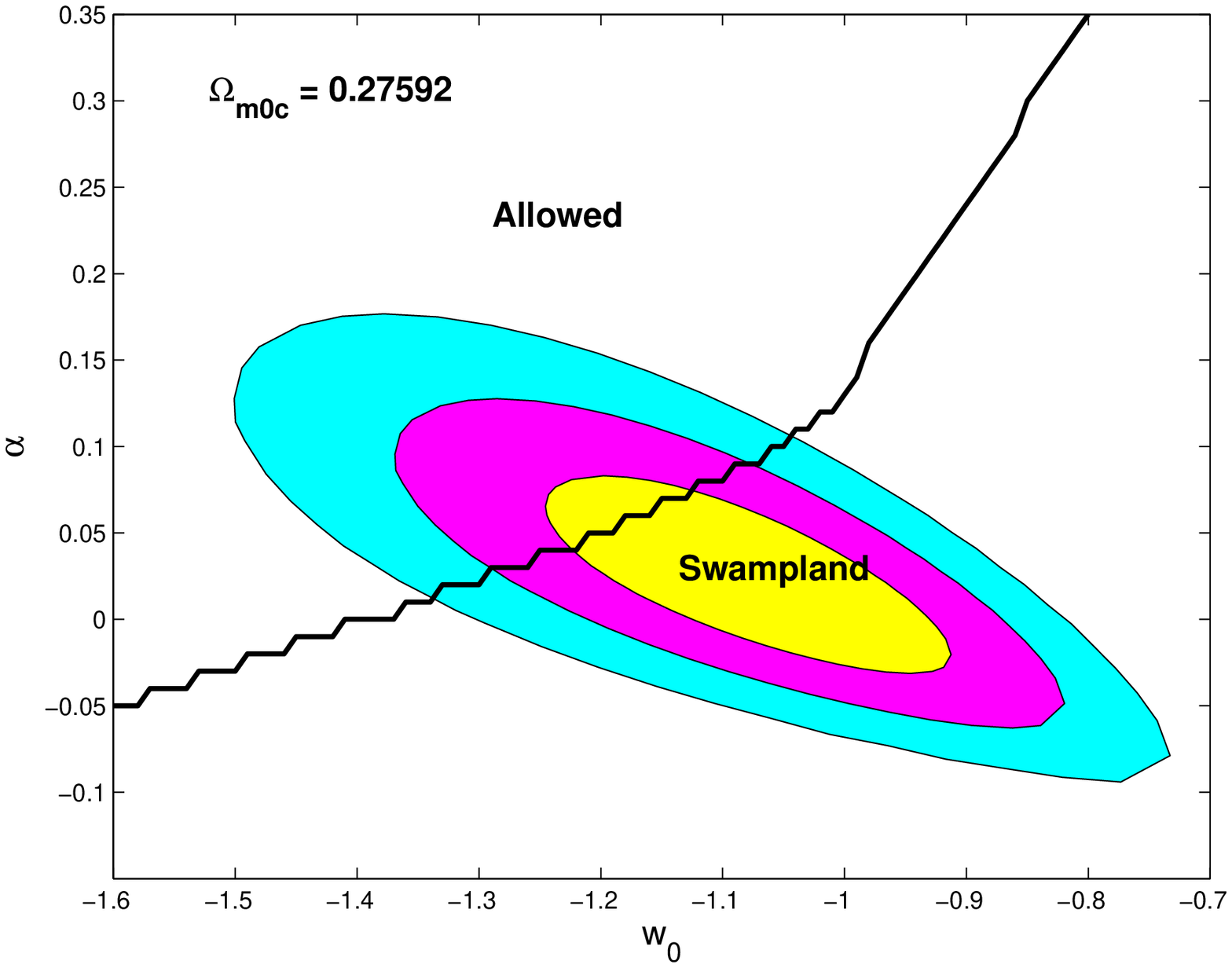}}
    \hspace{1in}
    \subfigure[]{
    \label{fig3:subfig:c}
    \includegraphics[width=6cm]{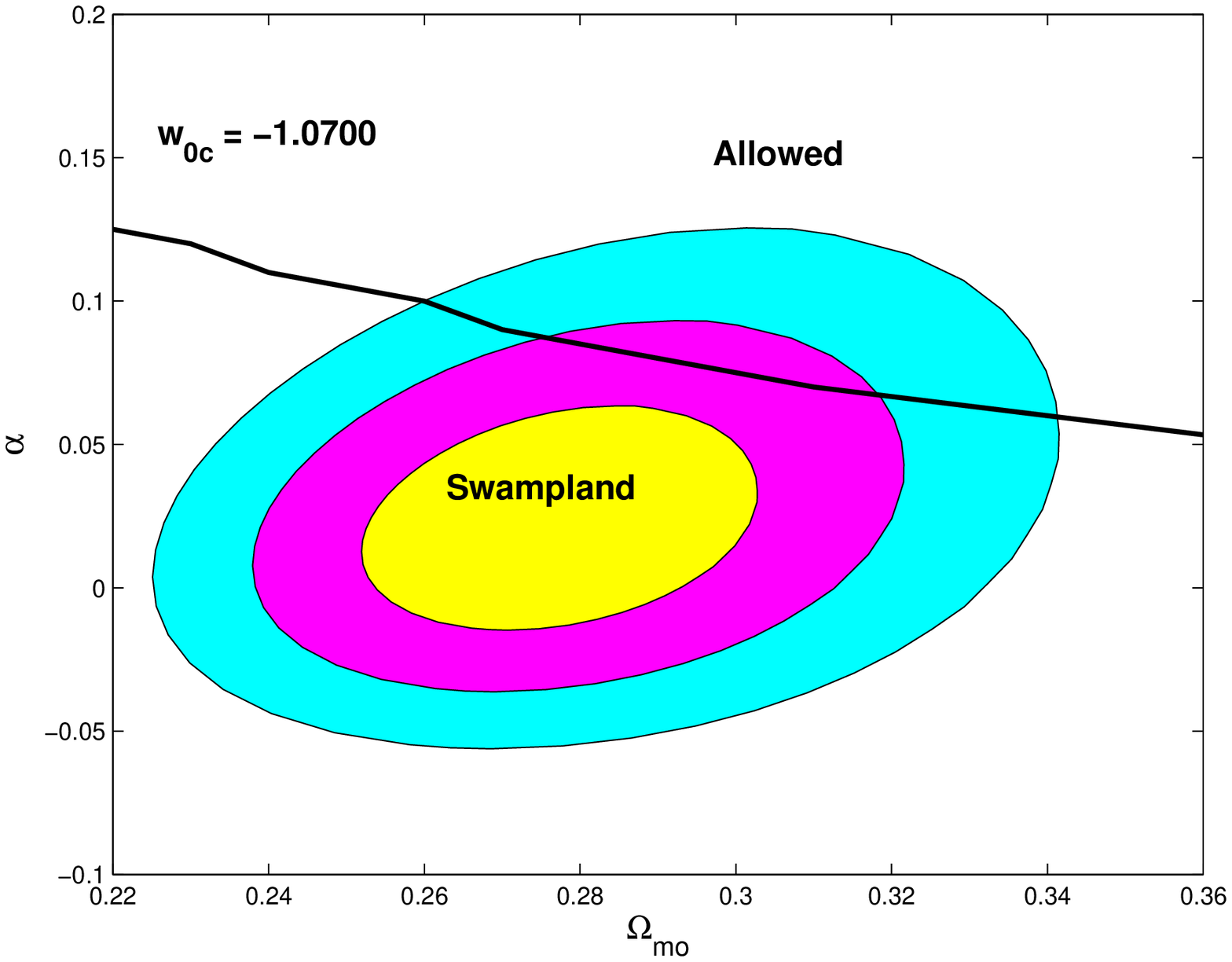}}
    \hspace{1in}
\caption{Contours on the parameter space from the observational fitting result by using the Union2 SNIa, the BAO measurement from SDSS and the
CMB shift parameter from WMAP5 for the interaction between dark sectors with the form $Q_1$ and time-dependent DE EoS
$w_2(z)=w_0\exp[z/(1+z)]/(1+z)$. The solid line indicates the constraint from the theoretical condition. ``allowed" indicates the region permitted
by the weak gravity conjecture. In (a) we choose the  $\alpha_c=0.02194$. In (b) we take $\Omega_{m0c}=0.27592$. In (c), $w_{0c}=-1.0700$. }
  \label{fig3}
\end{figure}

\begin{figure}
  \centering
  \subfigure[]{
    \label{fig4:subfig:a}
    \includegraphics[width=6cm]{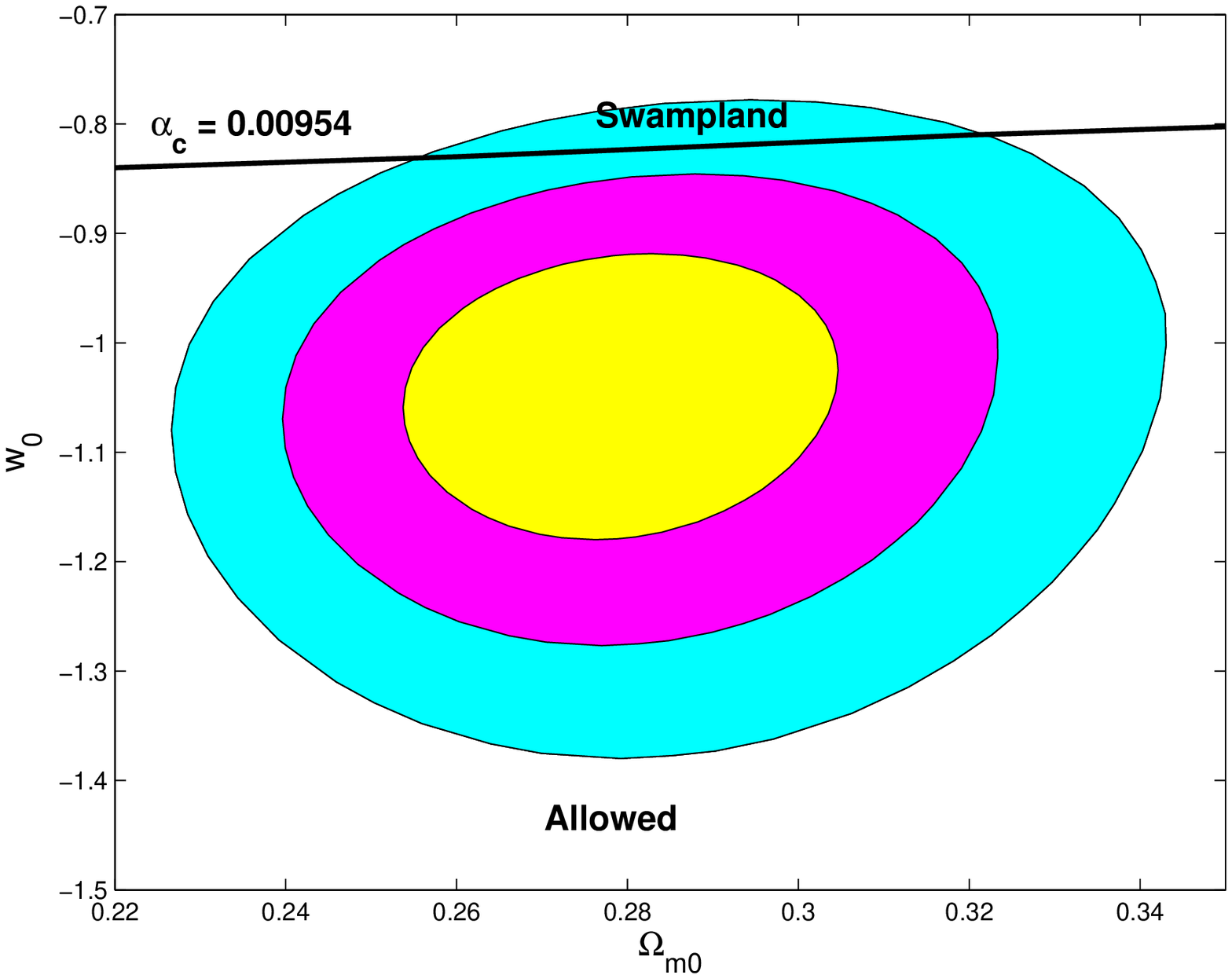}}
  \hspace{1in}
  \subfigure[]{
    \label{fig4:subfig:b}
    \includegraphics[width=6cm]{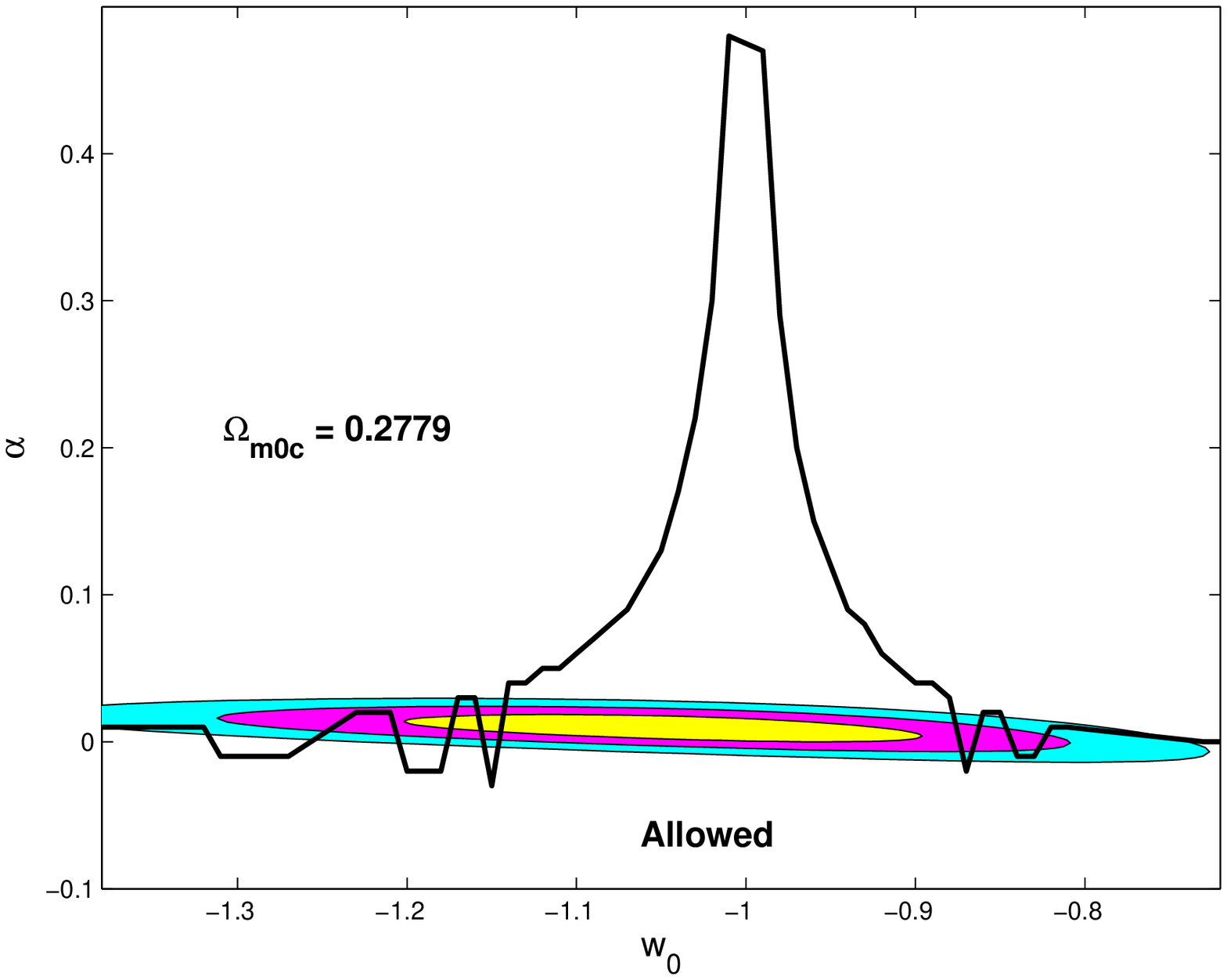}}
     \hspace{1in}
      \subfigure[]{
    \label{fig4:subfig:c}
    \includegraphics[width=6cm]{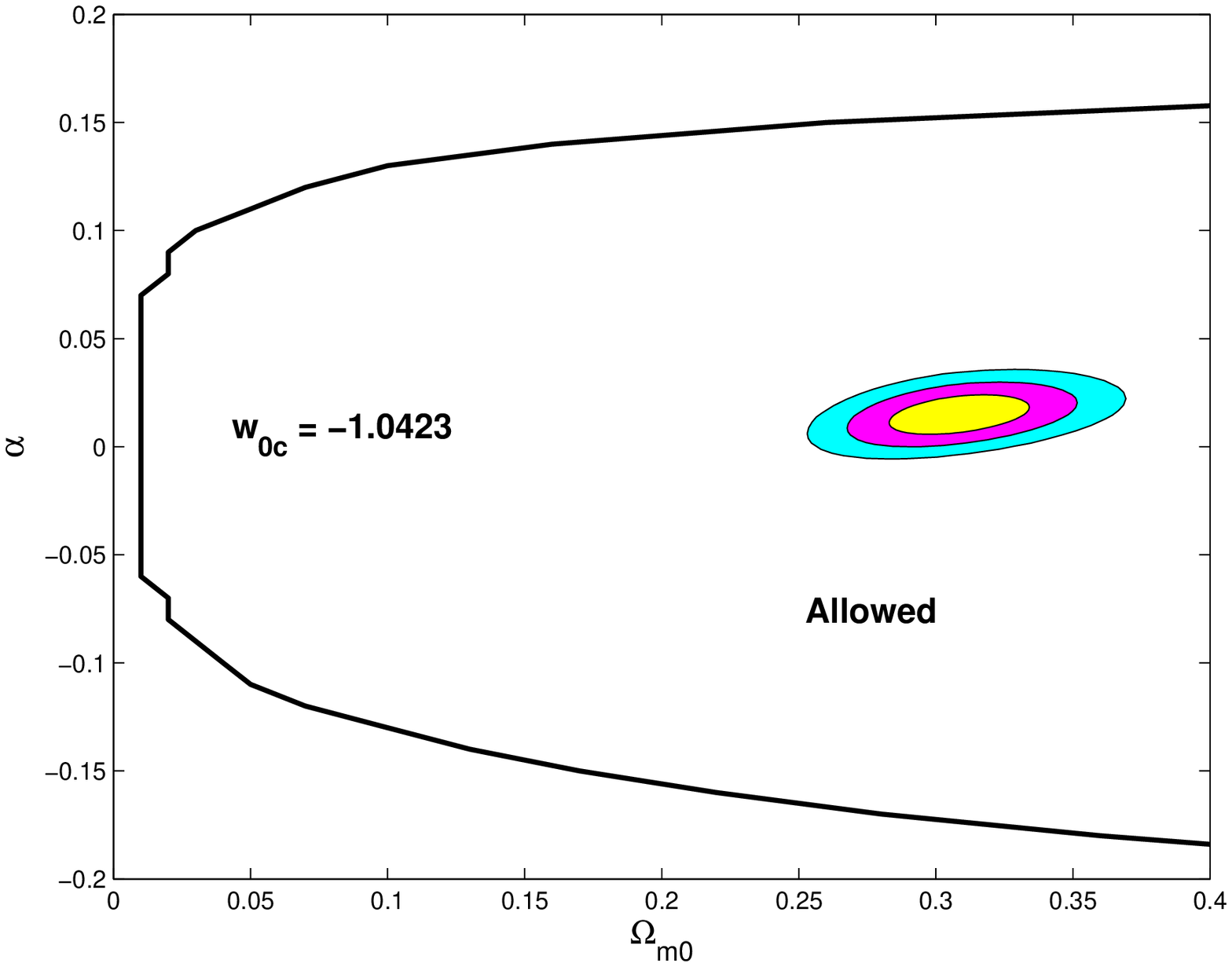}}
     \hspace{1in}
\caption{Contours on the parameter space from the observational fitting result by using the Union2 SNIa, the BAO measurement from SDSS and the
CMB shift parameter from WMAP5 for the interaction between dark sectors with the form $Q_2$ and constant DE EoS $w_1(z)=const$. The solid line
indicates the constraint from the theoretical condition. ``allowed" indicates the region permitted by the weak gravity conjecture. In (a) we
choose the  $\alpha_c=0.00954$. In (b) we take $\Omega_{m0c}=0.2779$. In (c), $w_{0c}=-1.0423$.}
  \label{fig4}
\end{figure}

\begin{figure}
\centering
\includegraphics[width=7cm]{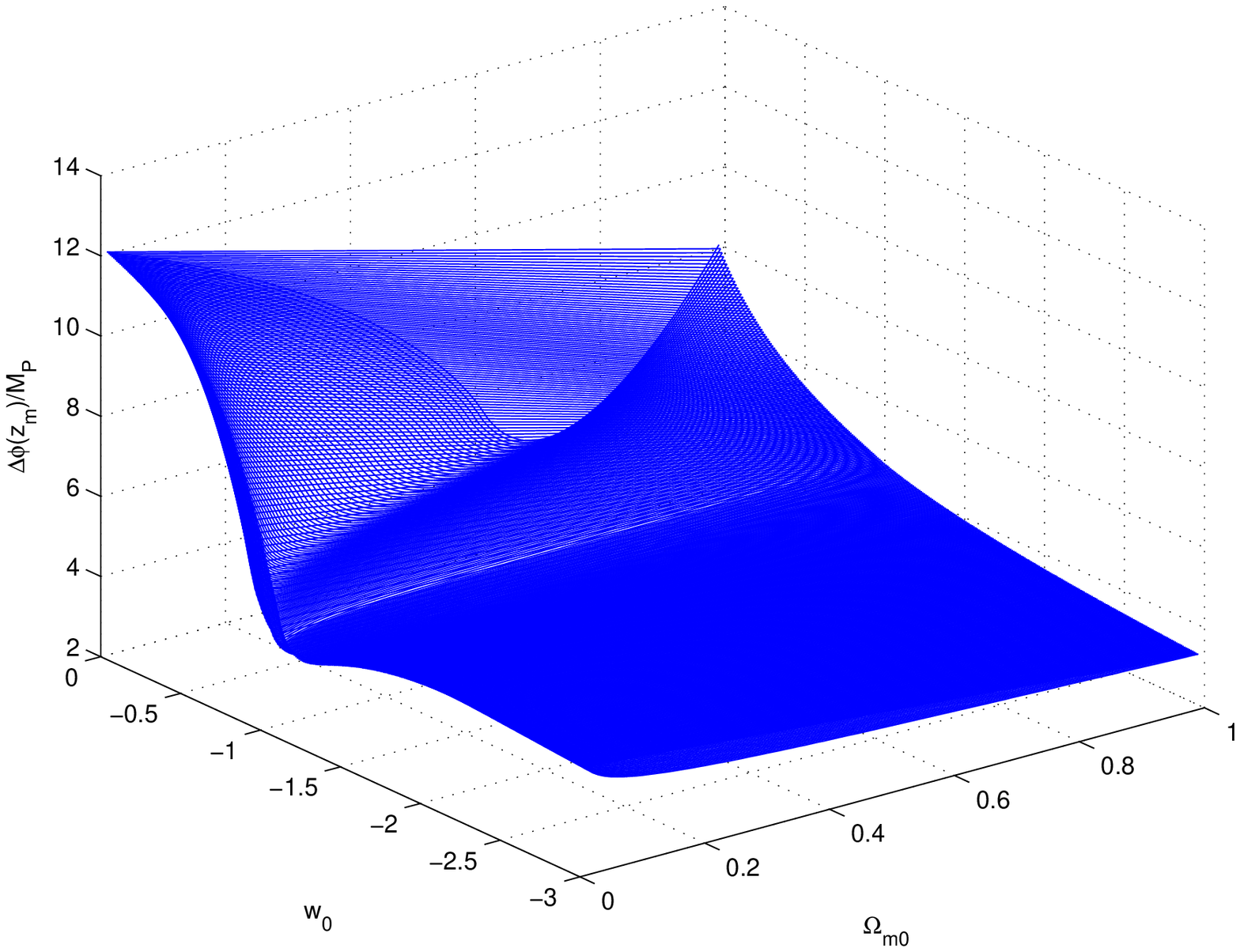}
\caption{$\Delta\phi{(z_{m})}/M_{P}$ is considerably larger than $1.00$ for different $\Omega_{m0}$, $w_0$ with $\alpha=0.030$ in the interacting
DE form $Q_2$ with $w_2(z)=w_0exp[z/(1+z)]/(1+z)$. } \label{fig5}
\end{figure}

\subsection{Interaction form 2}

Now we turn to the discussion of the interaction between dark sectors with the phenomenological form $Q=\alpha \kappa^{2n} H^{3-2n} \rho_m^n$. The dynamics
of DE and DM with this interaction was discussed in \cite{Chen:2008pz}. For simplicity in the following discussion we take $n=2$, so that
$Q_2=\alpha \kappa^{2} H^{-1} \rho_m^2$.

We will first examine the DE with constant EoS $w_1(z)=w_0=const$. By substituting $Q_2$ and $w_1(z)=w_0$ into (\ref{dotz2}), we have
\begin{equation}\label{dotz6}
\frac{d\Omega_{\phi}}{dz}=\frac{3(1-\Omega_{\phi})}{1+z}[w_0\Omega_{\phi}-\alpha(1-\Omega_{\phi})],
\end{equation}
Inserting the solution of (\ref{dotz6}) into (\ref{weakgravconj}), we can discuss the weak gravity conjecture. We have three free
parameters ($\Omega_{m0}$, $w_0$, $\alpha$). To satisfy the weak gravity conjecture, we require $0.03\le\Omega_{m0}\le0.40$ and
$-1.20\le{w_0}\le{-0.67}$ by fixing $\alpha_c=0.00954$, $0\le{\alpha}\le0.48$ and $-1.37\le{w_0}\le{-0.72}$ by fixing $\Omega_{m0c}=0.2779$, and
$0.01\le{\Omega_{m0}}\le0.44$ and $-0.05\le{\alpha}\le{0.16}$  by fixing $w_{0c}=-1.0423$.

The evolutions of DM and DE are described by
\begin{equation}\label{rhozeq3}
\frac{d\rho_m/\rho_{m0}}{dz}=\frac{3}{1+z}\frac{\rho_{m}}{\rho_{m0}}
+\frac{3\alpha}{1+z}\frac{\Omega_{m0}}{E^2(z)}(\frac{\rho_{m}}{\rho_{m0}})^2,
\end{equation}
\begin{equation}\label{rhozeq4}
\frac{d\rho_{\phi}/\rho_{\phi0}}{dz}=\frac{3[1+w_0]}{1+z}\frac{\rho_{\phi}}{\rho_{\phi0}}
-\frac{3\alpha}{1+z}\frac{\Omega^2_{m0}}{1-\Omega_{m0}}\frac{1}{E^2(z)}(\frac{\rho_{m}}{\rho_{m0}})^2.
\end{equation}
With these equations at hand, we can check the theoretical constraint from the weak gravity conjecture by comparing it with the observational
constraints. The best fit values to Union2 SNIa+BAO+$R$ data are $\Omega_{m0}= 0.278^{+0.074}_{-0.056}$,
$w_0=-1.042^{+0.385}_{-0.431}$ and
$\alpha=0.010^{+0.022}_{-0.026}$ with $\chi^{2}=547.16$ at the $3\sigma$ confidence level. The strength of the interaction is severely
constrained within the range $ -0.016\le{\alpha}\le0.032$ at the $3\sigma$ confidence level. If we use the
Constitution SNIa+BAO+$R$ data sets, the joint
constraints are $\Omega_{m0}= 0.281^{+0.071}_{-0.064}$, $w_0=-0.981^{+0.199}_{-0.263}$ and $\alpha= 0.008^{+0.026}_{-0.031}$ with
$\chi^{2}=466.41$ at the $3\sigma$ confidence level. The constraint on the strength of the interaction is a little better from the Union2 SNIa data.

We plot the contours of the observational constraints at $3\sigma$ confidence levels together with the result from the weak gravity conjecture in
Fig. 4, where we adopted $\alpha_c=0.00954$, $\Omega_{m0c}=0.2779$ and $w_{0c}=-1.0423$, respectively. Combining with the observation, in contours
$\Omega_{m0}$-$w_0$ and $\alpha$-$w_0$, we see that the theoretical constraint can help to reduce the parameter space. But the theoretical
constraint from the weak gravity conjecture cannot put stringent limit on the strength of the interaction if compared with the observational
constraint. This is shown in  Fig. \ref{fig4:subfig:c}. In Fig. \ref{fig2:subfig:b} we also plot $\Delta\phi/M_p$ versus $z_m$ for different
$\Omega_{m0}$, $w_0$ and $\alpha$ for the interacting dark energy form $Q_2$, where we see that the variation experienced by the DE scalar field
within the classical expansion era till now can not exceed Planck's mass.

We also examine the time-dependent DE EoS with the form $ w_2(z)=w_0\exp[z/(1+z)]/(1+z)$. Substituting  $ w_2(z)$  into (\ref{rhozeq4}), we can
constrain the model from observations. Fitting to the Union2 SNIa+BAO+$R$ data, we get the joint constraints on the parameters  $\Omega_{m0}=
0.276^{+0.073}_{-0.055}$, $w_0=-1.063^{+0.357}_{-0.433}$ and $\alpha= 0.007^{+0.004}_{-0.081}$ with $\chi^{2}=546.97$ at
the $1\sigma$ confidence level. If we use the Constitution SNIa+BAO+$R$ data sets, the joint constraints are $\Omega_{m0}= 0.280^{+0.030}_{-0.029}$,
$w_0=-0.995^{+0.095}_{-0.107}$ and $\alpha= 0.002^{+0.022}_{-0.029}$ with $\chi^{2}=466.40$ at the $1\sigma$ confidence level.

Inserting $ w_2(z)$  into (\ref{weakgravconj}) and substituting $ w_2(z)$ and $Q_2$ into (\ref{dotz2}), we can easily obtain the ratio between
the variation experienced by the DE scalar field and the Planck's mass
\begin{eqnarray}
\label{weakgravconj3}
\frac{\Delta\phi(z_m)}{M_p}=
\int_0^{z_m}\frac{\sqrt{3\mid[1+w_0exp(z/(1+z))/(1+z)]\Omega_{\phi}(z)\mid}}{1+z}dz,
\end{eqnarray}
where the evolution of the DE reads
\begin{equation}\label{dotz6b}
\frac{d\Omega_{\phi}}{dz}=\frac{3(1-\Omega_{\phi})}{1+z}[w_0\exp(z/(1+z))/(1+z)\Omega_{\phi}-\alpha(1-\Omega_{\phi})].
\end{equation}
When the strength of the interaction between dark sectors falls in the observational range $-0.006\le{\alpha}\le0.034$, we find that
$\Delta\phi(z_m)/M_p$ is considerably larger than $1.00$. In Fig. 5 we report this result by taking $\alpha=0.030$. This shows that for the
interaction between DE and DM with the form $Q_2$, if the DE EoS is time evolving as $ w_2(z)$, the weak gravity conjecture breaks down. This
tells us that this interaction model is not a viable model.

\section{Discussions}

In summary we have generalized the discussion on using the weak gravity conjecture in constraining the DE to the interacting DE models. We
examined two plausible forms of the interaction between dark sectors on phenomenological bases with constant and time-dependent DE EoS. By
comparing with the observational constraints, we found that although the constraint given by the weak gravity conjecture is consistent with the
observational results, in most cases the theoretical constraints are looser, except in some specific situations that stringent constraints can be
got by combining the theoretical and observational constraints. Thus in more general DE models, the weak gravity conjecture is not as powerful as
reported in \cite{pavon1}. Because the Union2 data contains more SNIa data, the observational constraint on the strength of the interaction
by the Union2 SNIa data is a little better than that by the Constitution SNIa data.

\begin{acknowledgments}
The work is supported by the National Natural Science Foundation of China key project under grants No. 10878001, 10935013, the Ministry of
Science and Technology of China national basic science Program (973 Project) under grant No. 2010CB833004, and the Natural Science Foundation
Project of CQ CSTC under grant No. 2009BA4050. NP was partially supported by the project A2008-58 of Chongqing University of Posts and
Telecommunications, and the NNSF of China under grant no 10947178.
\end{acknowledgments}

\end{document}